\documentclass[%
 aip,
 amsmath,amssymb,
 reprint,%
]{revtex4-1}

\usepackage{graphicx}
\usepackage{dcolumn}
\usepackage{bm}

\usepackage[utf8]{inputenc}
\usepackage[T1]{fontenc}
\usepackage{mathptmx}
\usepackage{etoolbox}

\usepackage{parskip}
\usepackage{verbatim}
\usepackage{amsmath}
\usepackage{geometry}
\global\long\def\d{\mbox{d}}
\def\order#1{\mathcal{O}\left(#1\right)}

\makeatletter

\usepackage{slashed}
\makeatother

\usepackage{babel}
\global\long\def\d{\mbox{d}}%

\makeatletter
\def\@email#1#2{%
 \endgroup
 \patchcmd{\titleblock@produce}
  {\frontmatter@RRAPformat}
  {\frontmatter@RRAPformat{\produce@RRAP{*#1\href{mailto:#2}{#2}}}\frontmatter@RRAPformat}
  {}{}
}%
\makeatother
\begin{document}


\title[Shock Wave in the Beirut Explosion: Theory and Video Analysis]{Shock Wave in the Beirut Explosion: Theory and Video Analysis}
\author{Adam J. Czarnecki}
\affiliation{Department of Physics, McGill University, Montreal, Quebec, Canada}

\author{Andrzej Czarnecki}%
\affiliation{Department of Physics, University of Alberta, Edmonton, Alberta, Canada}%
\email{andrzejc@ualberta.ca}

\author{Raquel Secrist}%
\affiliation{Department of Physics, University of Alberta, Edmonton, Alberta, Canada}%

\author{Julia Willsey}%
\affiliation{Department of Physics, University of Alberta, Edmonton, Alberta, Canada}%

\begin{abstract}
Videos of the 2020 Beirut explosion offer a rare opportunity to
see a shock wave. We summarize the non-linear theory of a weak shock,
derive the Landau-Whitham formula for the thickness of the
overpressure layer and, using frame-by-frame video analysis,
we demonstrate agreement of data and theory.
\end{abstract}

\maketitle

\section{Introduction}
On August 4, 2020, a catastrophic explosion of ammonium nitrate occurred at the Port of
Beirut, resulting in widespread destruction, over 200 fatalities, and
thousands of injuries \cite{Yammine2023}. 
Before the blast, a fire broke out in a harbor warehouse storing
fireworks, so many cameras were trained on the site. In addition, 
vehicles in Lebanon are often equipped with dash cameras. As a result,
many films surreptitiously captured the explosion and its blast
wave. They offer an opportunity to determine the dynamics of the wave.

Two regimes can be distinguished in the evolution of the blast
wave: strong and weak. The key parameter distingiushing them is the
ratio of the shock speed to the ambient sound speed, that is the Mach
number of the shock wave $M_{\rm sw}$. The dividing value can be taken
as \cite{WeiHargather2021}
 $M_{\rm sw}\simeq 5$.

 Immediately after the explosion follows a strong blast whose
propagation is influenced mainly by the inertia of ambient air and
not, for example, by pressure or temperature separately. 
In this regime, dimensional analysis relates the 
radius $R(t)$ of the shock front to the energy $E$ released in
the initial explosion at $t=0$,
\begin{equation}
R(t)=S(\gamma) \left(\frac{Et^2}{\rho_0} \right)^{1/5},
\label{Taylor}
\end{equation}
where $S(\gamma)$ is of the order of unity and depends on the ratio of
specific heats (adiabatic index)
$\gamma$; $\rho_0$ is the density of undisturbed air. 
This formula goes back to G.~I.~Taylor's famous estimate of the yield of the Trinity
Test\cite{GITaylor,GITaylor2}. 
Taylor's approach is carefully described in its historical
context in Ref.~[\!\!\citenum{Deakin2011}] and has recently been
revisited\cite{mone2024rev}.

More recently,  Wei and Hargather \cite{WeiHargather2021}
proposed a scaling that also includes the ambient sound speed $c_0$,
with characteristic length scale $l_c$ and scaled variables $R^\ast,t^\ast$,
\begin{equation}
l_c=\left(\frac{E}{\rho_0 c_0^2}\right)^{1/3},\qquad
R^*=\frac{R}{l_c},\qquad
t^*=\frac{t c_0}{l_c}.
\end{equation}
They demonstrated that the new scaling law collapses data for blasts
in air and in water to a single curve.
In the strong-shock regime the Mach number of the shock wave is so
much larger than the ambient sound speed, $M_{\rm sw}\gtrsim 5$, that
$c_0$ ceases to be a relevant control parameter and one recovers
$R^\ast \propto (t^\ast)^{2/5}$. As far as we
know, the new scaling law has not been applied to the Beirut
explosion yet. 

Published analyses are based on the Taylor-Sedov law, Eq.~\eqref{Taylor}. Rigby et
al.~\cite{Rigby:2020}~extracted blast-wave arrival times from social-media videos  and
fitted Eq.~\eqref{Taylor}, obtaining a best estimate 
of 0.5~kt TNT equivalent with an upper limit of
1.12~kt. Dewey~\cite{DeweyTNT} reanalyzed those arrival-time
data and showed that TNT and ammonium nitrate equivalence vary with distance from
the source rather than being a fixed number. Using near-field
fireball kinematics and Taylor-Sedov scaling, Aouad et
al.~\cite{Aouad:2021aa} obtained a lower estimate of about 0.2 kT TNT. 
Independent geophysical constraints were
supplied by Pilger et al.~\cite{Pilger:2021aa}, who combined
seismic, hydroacoustic, infrasonic, and radar-remote-sensing data, and
by Kim and Pasyanos~\cite{KimPasyanos2022Beirut}, who used
infrasound waveform inversion.

\begin{figure}
(a) \includegraphics[scale=0.9]{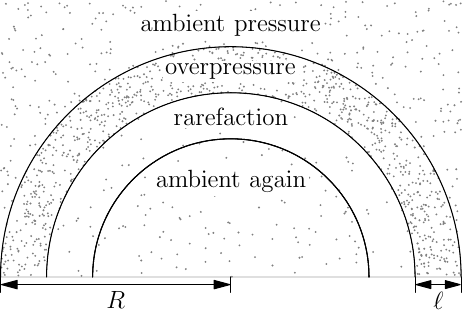} 

\vspace*{4mm}
(b) \includegraphics[scale=0.48]{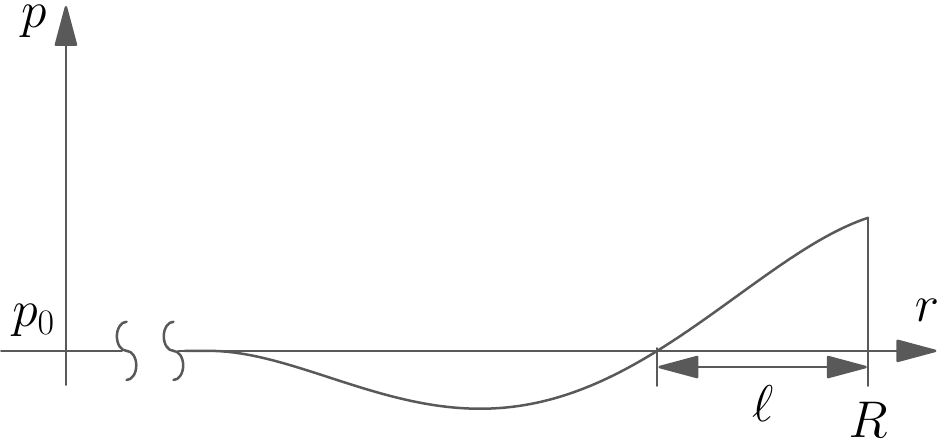} 

\caption{\label{fig:Shock} (a) Sketch of the structure of the blast 
  wave. The lower pressure in the rarefaction layer causes water 
  vapor to condense, resulting in the white cloud 
  visible behind the shock front.
(b) Dependence of pressure $p$ on the distance $r$ from the
explosion. The shock is followed by the overpressure region of width
$\ell$, and then by a region of decreased pressure. (Adopted from Ref.~[\!\!\citenum{Zeldovich46}]).} 
\end{figure}

As the blast propagates, it spreads and its overpressure becomes small
in comparison with the ambient pressure. This is the second, weak-shock
regime, and it is the primary focus of this paper. 

Fig.~\ref{fig:Shock} illustrates the structure of the shock wave: the
leading high-pressure region of thickness $\ell$ is followed by a region
of low pressure in which water vapor condenses, creating a white
so-called Wilson cloud, named after the 1927 Physics Nobel
laureate, C.~T.~R.~Wilson. Although from the outside it appears as if
the whole volume of 
the wave is opaque, the layer of condensed vapor is similarly thin as
the high pressure layer. Video recordings show
vapor clouds quickly vanishing after the shock passes.

Z\'ehil \cite{Zehil:2024} analyzed the
interaction of the blast wave with the Beirut port silos and concluded that
the apparent window in the Wilson cloud  arises from
thermodynamic conditions for condensation behind the shock.  

In some of the footage of the Beirut explosion, the
high-pressure region, having a larger density and therefore larger
refraction index than ambient air, is visible in front of the
condensation cloud. Thus we can determine the growth of its thickness with
the distance covered. This growth is a non-linear phenomenon, unlike
for example propagation of sound, based on a linear theory.

An excellent textbook\cite{FaberForPhysicists} states that {\it shock
  fronts are not visible in the world about us in the way that tidal
  bores and hydraulic jumps are}, contrasting the harder to observe
aerodynamic phenomena with spectacular effects in liquids. However,
thanks to ubiquitous cameras, shock fronts from accidental explosions
have been recorded and can be examined.

Hints of shock-like behavior in solutions of the Euler equations were
noted already in the mid-19th century -- for example by James Challis in
1848 -- although their physical significance was not recognized until
later developments by Riemann and by Rankine and Hugoniot (see Section
\ref{sec:RH}). An excellent
historical account of the early development of shock-wave theory is
given in
Ref.~[\!\!\citenum{Salas2007TheCE}]). 

Subsequent progress was driven largely by physicists working on
compressible hydrodynamics in the context of explosions, motivated in part by the
analysis of nuclear explosions.  These ideas were later systematized in
the textbook treatments of Zeldovich and Raizer \cite{zel2012shock}, in Landau and
Lifshitz's course \cite{LL6}, and most recently by David Tong
\cite{Tong2025fluid}.  Approximations for  nonuniform flows were developed by Whitham and
others using geometric shock dynamics \cite{emanuel2022three,Best1991GSDReview,whitham:1974}, while Lighthill
clarified the nonlinear steepening of finite-amplitude waves \cite{lighthill1978book}.  Today shock-wave
theory continues to find applications across physics, ranging from
astrophysical blast waves to relativistic hydrodynamics and heavy-ion
collisions \cite{DerradideSouza:2015kpt,Rocha:2023ilf}. 

For weak shocks, the theory of the asymptotic evolution
of the overpressure layer was pioneered independently  by Landau\cite{Landau1945},
Bethe et al.~\cite{bethe1958blast}, and
by  Whitham\cite{Whitham:1950, whitham_1956},
who showed that as $t\rightarrow \infty$, the layer’s
thickness $\ell$ is related to $R$ by 
\begin{equation} \label{main}
\ell(t) \propto \sqrt{\ln \frac{R(t)}{a}},
\end{equation}
where $a$ is a constant dependent on short-distance details of the
explosion. To our knowledge, no single textbook or paper contains a
clear derivation of this equation. Existing discussions are either
cryptic or rely on unrealistic assumptions. In Section  \ref{theory}, we provide
a derivation that draws on several sources.

In Section \ref{video} we analyze publicly available video footage of the
Beirut explosion to determine the  evolution of the blast
wave’s structure. By measuring the apparent thickness of the
overpressure region at various distances, we assess the
applicability of the weak-shock scaling in Eq.~\eqref{main}. We note that the
predicted square root of a logarithm is a very slowly increasing
function. Given the scarcity of data and far from perfect quality of
the videos we do not claim high precision. However, we do find
evidence of the increase of thickness and are able to fit 
the prediction of Eq.~\eqref{main} to data. 

Section \ref{sec:pedagogy} summarizes the pedagogical value of this
study and outlines possible follow-up student projects.

\section{Theory of a weak spherical blast}
\label{theory}

If an explosion can be approximated as point-like, it launches a
spherically symmetric blast wave. At time $t$ from the explosion, the
shock front is at the radial distance $R(t)$ and moves outward with
speed $D=\d R/\d t$. Immediately behind the shock the pressure $p^\ast$
exceeds ambient $p_0$ (asterisk marks quantities referring to
gas immediately following the shock and subscript 0 describes
undisturbed conditions). We call the shock weak if the fractional
overpressure,
\begin{equation}
\pi^\ast(R)\equiv \frac{p^\ast(R)-p_0}{p_0},
\end{equation}
is small,  $\pi^\ast\ll 1$. 
In that regime, one can predict the radial thickness $\ell$ of the
overpressure region.
The goal of this section is to derive the large-$R$ behavior,
\begin{equation}
  \label{eq:1}
  \ell(R) \propto \sqrt{\ln(R/R_0)} \quad (R\to \infty),
\end{equation}
where $R_0$ is a reference radius introduced to make the argument
dimensionless, and the proportionality constant and asymptotically
smaller terms depend on the short-distance properties of the explosion
(the early-time waveform). 
 
\subsection{Rankine-Hugoniot jump conditions}
\label{sec:RH}
The shock is a thin layer across which the air density $\rho$, its
velocity $u$, and pressure $p$ jump. Jump conditions were formulated
in 19th century by Rankine and by Hugoniot for a planar shock.  For a spherical shock at large
$R$, these ``RH'' conditions are the same up to effects of higher order in
$1/R$ than our objective.

Work in the shock frame.
The unshocked air is in the state $(\rho_0,p_0)$ and flows into the shock with speed $D$.
The state just behind the shock is $(\rho^\ast,p^\ast)$; in the shock frame it flows
out with speed $D-u^\ast$.
Mass, momentum, and energy conservation across the discontinuity give
\begin{align}
\rho_0 D &= \rho^\ast(D-u^\ast), & \text{RH1}
\label{eq:RH_mass}
\\
 p_0+\rho_0 D^2 &= p^\ast+\rho^\ast(D-u^\ast)^2,  & \text{RH2}
\label{eq:RH_mom}
\\
 e_0+{p_0\over \rho_0}+ \tfrac12 D^2 &= e^\ast+{p^\ast\over
                                       \rho^\ast}+\tfrac12
                                       (D-u^\ast)^2, & \text{RH3}
\label{eq:RH_energy}
\end{align}
where $e$ is the internal energy per unit mass of air.
We describe air as an ideal gas. 
Then $e+p/\rho =\gamma p/[(\gamma-1)\rho]$. (The left hand side is 
enthalpy per unit mass; the equality follows from the ideal gas law $p=\rho R_uT/M$, $e=c_v T$, $c_p/c_v =
\gamma$,  and $c_p = c_v + R_u/M$,
where $R_u$ is the universal gas constant and $M$ is the molar mass of
air. $c_{p,v}$ denote specific heats per unit mass.)

For a compressive shock the momentum condition says that the gas
slows down while the pressure rises. The energy condition implies that
part of the kinetic energy is converted into internal energy by
the work done during compression. 

Crucially for our purposes, the entropy increase is of the third order
of smallness, if the pressure and density changes are counted as first
order. 
This smallness of entropy production was first established
by Jouguet\cite{ShockDecade} and is explained in the following section.

\subsection{Entropy production is negligible}
\label{app:entropy}
In an ideal gas the change of entropy per unit mass, $\Delta s$, is
obtained from the first law of thermodynamics and the ideal gas law,
\begin{equation}
  \label{eq:2}
  {\Delta s \over c_v} = \Delta \ln {p \over \rho^\gamma}.
\end{equation}
We will use the jump conditions alone to show that this change across
the shock is $\order{\pi^{\ast 3}}$.  

We measure disturbances relative to the undisturbed atmosphere,
\begin{equation}
\delta \equiv \frac{\rho-\rho_0}{\rho_0},\qquad
\pi \equiv \frac{p-p_0}{p_0},\qquad
c_0=\sqrt{\gamma p_0/\rho_0}.
\label{eq:Distur}
\end{equation}
In the time window analyzed below in Sec.~\ref{video},  $1.93<t<3.17$~s, the shock speed is $D\simeq365$ m/s. 
Using the atmospheric values employed in Sec.~\ref{sec:hum},
$p_{0}\simeq101$ kPa and $\rho_{0}=1.2 \text{ kg/m}^{3}$, the ambient sound speed
in $c_{0}\simeq343$ m/s. Thus the Mach number of the shock is
$D/c_{0}\simeq1.06$, far below the strong-shock regime \cite{WeiHargather2021} $M_{{\rm
    sw}}\gtrsim5$. Via Eq.~\eqref{eq:D_weak}, this
implies $\pi,\delta$, and $u/c_{0}$ are all at most of order
$10^{-1}$, which justifies applying weak-shock asymptotics to the
far-field portion of the blast wave.

Recall our  convention of using the asterisk for quantities
immediately behind the shock,
$\rho^\ast /\rho_0 = 1+\delta^\ast$, $p^\ast /p_0 = 1+\pi^\ast$. From
RH1,
\begin{equation}
  \label{eq:3}
  D-u^\ast = {D\over 1+\delta^\ast}.
\end{equation}
Substitute this expression into RH2 and denote by $c_0$ the sound
speed in undisturbed air, $c_0^2 = \gamma p_0/\rho_0$, \
\begin{equation}
  \label{eq:4}
  \pi^\ast = {\gamma\delta^\ast \over 1+\delta^\ast} {D^2 \over c_0^2}.
\end{equation}
The same substitution converts RH3 into
\begin{equation}
  \label{eq:5}
  {2 \over \gamma-1}{\pi^\ast - \delta^\ast \over 1+\delta^\ast} =
  \left[ 1 - {1\over (1+\delta^\ast)^2}\right] {D^2 \over c_0^2}.
\end{equation}
Eliminating $D/c_0$ from the last two equations we find an exact
relationship between pressure and density disturbances caused by the
shock,
\begin{equation}
  \label{eq:6}
  \delta^\ast = {2\pi^\ast \over 2\gamma +(\gamma-1)\pi^\ast}.
\end{equation}
We use it to determine $\Delta s/c_v$ in Eq.~\eqref{eq:2},
\begin{align}
  \ln {p^\ast \over \rho^{\ast \gamma}}
 -  \ln {p_0 \over    \rho_0^\gamma} 
&=  \ln  {1+\pi^\ast \over (1+\delta^\ast)^\gamma}\\
&= \frac{\gamma ^2-1} {12 \gamma ^2}
   \pi^{\ast 3}
 + \order{\pi^{\ast 4}}.
 \label{eq:7}
\end{align}
Since $\gamma > 1$, the entropy is slightly increased by the shock. 
We are interested at most in second order effects in the
discussion that follows and can treat the weak shock
as an adiabatic process. 

\subsection{Governing equations outside the shock}

Outside the thin shock layer, mass
and momentum conservation are expressed by the continuity 
and Euler equations. For a spherically symmetric flow with purely
radial velocity $u(r,t)$,
\begin{align}
\partial_t \rho
 + \frac{1}{r^2}\,\partial_r\!\left(r^2\rho  u\right)
 &=0\text{ (continuity)},
\label{eq:cont_sph}\\
\partial_t u + u\,\partial_r u+\frac{1}{\rho}\,\partial_r p
 &=0 \text{ (Euler)}.
\label{eq:euler_sph}
\end{align}
To close these equations we use the adiabatic condition
$p\rho^{-\gamma}=\text{const}$ (Sec.~\ref{app:entropy}). 
The sound speed is
\begin{equation}
c^2\equiv\left(\frac{\partial p}{\partial \rho}\right)_{s}
=\gamma\,\frac{p}{\rho}.
\label{eq:def_c}
\end{equation}
Subscript $s$ indicates that the derivative is taken at constant
entropy per unit mass.

\subsection{Linear spherical acoustics: $\pi\propto 1/r$ and $u\simeq c_0\pi/\gamma$}
\label{sec:linear_acoustics}

Two large-$r$ facts will be used repeatedly:
(i) the overpressure amplitude decays as $1/r$;
(ii) for an outgoing weak wave, the velocity satisfies
$u\simeq c_0\pi/\gamma$.
We derive both by linearizing Eqs.~\eqref{eq:cont_sph}-\eqref{eq:euler_sph},
\begin{align}
\partial_t \delta +  \frac{1}{r^2}\,\partial_r\!\left(r^2 u\right) &= 0,
\label{eq:cont_lin}
\\
\partial_t u + {c_0^2\over \gamma} \partial_r \pi &= 0.
\label{eq:euler_lin}
\end{align}
For adiabatic disturbances
$\delta = \pi /\gamma $. 
Eliminating $\partial_t\partial_r(r^2 u)$ yields the spherical wave equation
\begin{equation}
\partial_t^2\!\left(r\,\pi\right)=c_0^2\,\partial_r^2\!\left(r\,\pi\right).
\label{eq:spherical_wave}
\end{equation}
Equation~\eqref{eq:spherical_wave} has the general solution
\begin{equation}
r\pi(r,t)=F\!\left(t-\frac{r}{c_0}\right)+G\!\left(t+\frac{r}{c_0}\right).
\label{eq:outgoing_incoming}
\end{equation}
The function $F$ describes a wave traveling outward, while $G$
describes one traveling inward. 
For a blast launched at the origin we keep only the outward-moving part,
\begin{equation}
\pi(r,t)=\frac{1}{r}F\!\left(t-\frac{r}{c_0}\right).
\label{eq:pi_outgoing}
\end{equation}
This is the desired $1/r$ decay of the amplitude.

We next derive the relation between $u$ and $\pi$ in the same linear
regime. Eq.~\eqref{eq:euler_lin} gives
\begin{equation}
\partial_t u=-{c_0^2 \over \gamma}\,\partial_r\pi.
\label{eq:euler_lin_pi}
\end{equation}
Substituting Eq.~\eqref{eq:pi_outgoing} and neglecting terms suppressed
by $1/r^2$,
\begin{equation}
\partial_t u={c_0 \over \gamma}\,\partial_t \pi,
\label{eq:u_pi_linear}
\end{equation}
leads to the desired relation 
\begin{equation}
  \label{eq:8}
u=c_0 \pi/\gamma.  
\end{equation}
\subsection{Propagation speed of a weak outward disturbance}
\label{sec:prop_speed}

The width $\ell$ grows because a compressive part of the wave
propagates slightly faster than $c_0$. Indeed, if the air is moving
with speed $u$ with respect to the ground and sound propagates with
speed $c$ with respect to air, we expect the speed of small
disturbances to be $c\pm u$. 
We now derive that speed from the governing equations.

Rewrite the continuity equation as
\begin{equation}
\partial_t \rho+\partial_r(\rho u)+\frac{2\rho u}{r}=0.
\label{eq:cont_split}
\end{equation}
The last term contains no radial derivatives. It acts as a source that influences amplitudes, but not the 
transport speeds. Thus, to determine how a short wave feature moves through a slowly varying flow,
we keep only the part containing derivatives (the so-called principal
part of the partial differential equation),
\begin{align}
\partial_t \rho +u\,\partial_r\rho+\rho\,\partial_r u &=0,
\label{eq:planar_cont}
\\
\partial_t u+u\,\partial_r u+\frac{1}{\rho}\,\partial_r p&=0.
\label{eq:planar_euler}
\end{align}

At a given point of the flow, regard $(\rho,u,p)$ as locally constant and superpose a much smaller
disturbance:
\begin{equation}
\rho \to \rho+\delta\rho,\qquad
u \to u+\delta u,\qquad
p \to p+\delta p.
\end{equation}
We keep only terms linear in $\delta\rho$, $\delta u$, and $\delta p$.

For adiabatic motion, $\delta p/p=\gamma\delta\rho/\rho$,
hence
\begin{equation}
\delta p=\frac{\gamma p}{\rho}\,\delta\rho=c^2\,\delta\rho,
\label{eq:delta_p}
\end{equation}
where $c^2=\gamma p/\rho$ is the local sound speed.

Linearizing Eqs.~\eqref{eq:planar_cont} and \eqref{eq:planar_euler} therefore gives
\begin{align}
(\partial_t+u\partial_r)\,\delta\rho + \rho\,\partial_r\delta u &=0,
\label{eq:delta_cont}
\\
(\partial_t+u\partial_r)\,\delta u + \frac{c^2}{\rho}\,\partial_r\delta\rho &=0.
\label{eq:delta_euler}
\end{align}
Since the background quantities $\rho$, $u$, and $c$ are treated as constants in this local
calculation, derivatives commute with them.  Apply $(\partial_t+u\partial_r)$ to
Eq.~\eqref{eq:delta_cont}:
\begin{equation}
(\partial_t+u\partial_r)^2\delta\rho
+\rho\,\partial_r\!\left[(\partial_t+u\partial_r)\delta u\right]=0.
\end{equation}
Now eliminate $(\partial_t+u\partial_r)\delta u$ with Eq.~\eqref{eq:delta_euler}. This yields
\begin{equation}
\left[(\partial_t+u\partial_r)^2-c^2\partial_r^2\right]\delta\rho=0.
\label{eq:convected_wave}
\end{equation}
Because $u$ and $c$ are constants here, the operator factors,
\begin{equation}
\left[\partial_t+(u-c)\partial_r\right]
\left[\partial_t+(u+c)\partial_r\right]\delta\rho=0.
\label{eq:factorized}
\end{equation}
Thus a small disturbance has two local characteristic velocities,
\begin{equation}
u-c
\qquad\text{and}\qquad
u+c.
\end{equation}
In the weak blast,
$u-c$ is negative and represents an inward-propagating disturbance, while 
$u+c$ is the outward-propagating one relevant here.
For the outward-moving part of the blast,
\begin{equation}
\frac{dr}{dt}=u+c.
\label{eq:uc_speed}
\end{equation}

We now evaluate $u+c$.  Eq.~\eqref{eq:8} gives,
\begin{equation}
u=\frac{c_0}{\gamma}\,\pi + \order{\pi^2}.
\label{eq:u_pi_linear_repeat}
\end{equation}
To express $c$ in terms of $\pi$, use the adiabatic relation
$p\rho^{-\gamma}=p_0\rho_0^{-\gamma}$ together with $p=p_0(1+\pi)$. Then
\begin{equation}
\frac{\rho}{\rho_0}
=\left(\frac{p}{p_0}\right)^{1/\gamma}
=1+\frac{\pi}{\gamma}+\order{\pi^2}.
\label{eq:rho_pi_expand}
\end{equation}
Therefore
\begin{equation}
\frac{c^2}{c_0^2}
=\frac{\gamma p/\rho}{\gamma p_0/\rho_0}
=\frac{1+\pi}{1+\pi/\gamma+\order{\pi^2}}
=1+\frac{\gamma-1}{\gamma}\pi+\order{\pi^2},
\end{equation}
so
\begin{equation}
c=c_0\left(1+\frac{\gamma-1}{2\gamma}\pi\right)+\order{\pi^2}.
\label{eq:c_pi}
\end{equation}
Combining Eqs.~\eqref{eq:u_pi_linear_repeat} and \eqref{eq:c_pi}, we obtain
\begin{equation}
u+c
=c_0\left(1+\frac{\gamma+1}{2\gamma}\pi\right)+\order{\pi^2}.
\label{eq:uc_pi}
\end{equation}
We see that this propagation speed of a weak perturbation exceeds
$c_0$ by a small quantity of first-order.

\subsection{Weak-shock speed from the jump conditions}
\label{sec:weak_shock_speed}

Equations~\eqref{eq:4} and \eqref{eq:6} already contain the shock
speed $D$ in terms of $\pi^*$. 
From Eq.~\eqref{eq:6},
\begin{equation}
1+\delta^\ast = \frac{2\gamma +(\gamma +1)\pi^\ast}{2\gamma +(\gamma -1)\pi^\ast},
\end{equation}
so
\begin{equation}
\frac{1+\delta^\ast}{\delta^\ast}=\frac{2\gamma +(\gamma +1)\pi^\ast}{2\pi^\ast}.
\end{equation}
Substituting this into Eq.~\eqref{eq:4} gives the exact relation
\begin{equation}
\frac{D^2}{c_0^2}=1+\frac{\gamma+1}{2\gamma}\pi^\ast.
\label{eq:D_exact_from_pi}
\end{equation}
For a weak shock, $\pi^\ast\ll 1$, so taking the square root yields
\begin{equation}
D=c_0\left(1+\frac{\gamma+1}{4\gamma}\pi^\ast\right)+\order{\pi^{\ast 2}}.
\label{eq:D_weak}
\end{equation}

\subsection{Rays and the width of the overpressure pulse}
\label{sec:rays_width}

To describe the broadening of the overpressure region, we compare the
shock position $R(t)$ with the trailing edge of 
the region where the pressure is above ambient. Let $r_0(t)$ be defined by
\begin{equation}
p\bigl(r_0(t),t\bigr)=p_0.
\end{equation}
The width of the overpressure region is therefore $R(t)-r_0(t)$.

Near the trailing edge the disturbance is particularly weak. There
$\pi=0$, and Eq.~\eqref{eq:8} then gives 
$u=0$ to leading order. Thus the trailing edge moves, to the same
level of accuracy, with the ambient 
sound speed $c_0$. This implies
\begin{equation}
r_0(t)=c_0t+\order{1}
\qquad (t\to\infty).
\label{eq:r0_ct_order1}
\end{equation}
The bounded $\order{1}$ term reflects details of the early stage of the explosion. It does not affect the unbounded growth derived below.

It is therefore convenient to define
\begin{equation}
c_0t=R-\ell(R).
\label{eq:l_def}
\end{equation}
In other words, $\ell(R)$ is the distance by which the shock is ahead
of a sound signal that would have traveled outward at speed $c_0$. By
Eq.~\eqref{eq:r0_ct_order1}, this $\ell(R)$ differs from the physical
pulse width only by $\order{1}$ at large $R$.

To determine how $\ell$ grows, we mentally mark and follow a short part of the
outward-going waveform as it moves.  The trajectory of  that marked
part in the $(r,t)$ plane will be called a ray. In the linear
spherical solution, Eq.~\eqref{eq:pi_outgoing}, one has
\begin{equation}
\pi(r,t)=\frac{1}{r}\,F\!\left(t-\frac{r}{c_0}\right).
\end{equation}
A marked part of the waveform corresponds to a fixed value of the retarded time
$t-r/c_0$. Along its path,
\begin{equation}
r\pi=\text{const}.
\end{equation}

In the far field the amplitude still decreases mainly because of this
spherical spreading. We therefore keep the linear relation
$r\pi=\text{const}$ when following a ray, but use the first nonlinear
correction in the propagation speed. Corrections to the amplitude law
are of higher order in the small disturbance; they change only bounded
terms and do not alter the leading growth of $\ell$. This suggests
introducing
\begin{equation}
z\equiv \frac{\gamma+1}{2\gamma}\,r\pi.
\label{eq:z_def}
\end{equation}
For the purely linear solution, $z$ is constant along each ray. Using
Eq.~\eqref{eq:uc_pi}, the ray equation becomes
\begin{equation}
\frac{dr}{dt}=c_0\left(1+\frac{z}{r}\right).
\label{eq:ray_ode}
\end{equation}
This differential equation can be integrated along each ray. Choose a fixed matching radius
$R_m$ in the weak-shock region, and let $t_m(z)$ denote the time at which the ray labeled by $z$
passes through $r=R_m$. Then
\begin{equation}
c_0\bigl[t-t_m(z)\bigr]
=
r-R_m-z\ln\frac{r+z}{R_m+z}.
\label{eq:ray_int_law}
\end{equation}
We assume that the waveform at $R_m$ is smooth and monotone near the trailing edge, so $t_m(z)$ is
a $C^1$ function of $z$ there. Since the pulse amplitude at $R_m$ is finite, the relevant values of
$z$ lie in a bounded interval.

At the shock, $r=R$ and $\pi=\pi^\ast$, so Eq.~\eqref{eq:z_def} gives
\begin{equation}
z(R)=\frac{\gamma+1}{2\gamma}\,R\pi^\ast(R).
\label{eq:z_shock}
\end{equation}
Using Eq.~\eqref{eq:l_def}, namely $c_0t=R-\ell(R)$, Eq.~\eqref{eq:ray_int_law} yields
\begin{equation}
\ell(R)=R_m-c_0 t_m(z)+z\ln\frac{R+z}{R_m+z}.
\label{eq:l_with_C}
\end{equation}
Since $z/R=(\gamma+1)\pi^\ast/(2\gamma)\ll 1$, one may also write
\begin{equation}
\ell(R)=R_m-c_0 t_m(z)+z\ln\frac{R}{R_m+z}+\order{\frac{z^2}{R}}.
\label{eq:l_z_leading}
\end{equation}
We do not yet discard the bounded $z$-dependent terms; we first determine $z(R)$.

We now relate $d\ell/dR$ to $z(R)$.
Differentiate Eq.~\eqref{eq:l_def} with respect to $t$:
\begin{equation}
c_0 = D\left(1-\frac{d\ell}{dR}\right),
\end{equation}
so
\begin{equation}
\frac{d\ell}{dR}=1-\frac{c_0}{D}.
\label{eq:lprime_from_D}
\end{equation}
Substituting the weak-shock expansion \eqref{eq:D_weak} gives
\begin{equation}
\frac{d\ell}{dR}=\frac{\gamma+1}{4\gamma}\pi^\ast+\order{\pi^{\ast 2}}.
\end{equation}
Now use Eq.~\eqref{eq:z_shock} to eliminate $\pi^\ast$:
\begin{equation}
\frac{d\ell}{dR}=\frac{z(R)}{2R}+\order{\frac{z^2}{R^2}}.
\label{eq:lprime_z}
\end{equation}

Differentiating Eq.~\eqref{eq:l_with_C} gives
\begin{align}
\frac{d\ell}{dR}
&=
\frac{z}{R+z}
\\
&+
\frac{dz}{dR}
\left[
\ln\frac{R+z}{R_m+z}
-c_0 t_m'(z)
+\frac{z}{R+z}
-\frac{z}{R_m+z}
\right].
\label{eq:lprime_full}
\end{align}
Combining Eqs.~\eqref{eq:lprime_z} and \eqref{eq:lprime_full}, and using
\begin{equation}
\frac{z}{R+z}=\frac{z}{R}+\order{\frac{z^2}{R^2}},
\end{equation}
we obtain
\begin{align}
\frac{dz}{dR}
&\left[
\ln\frac{R+z}{R_m+z}
-c_0 t_m'(z)
+\frac{z}{R+z}
-\frac{z}{R_m+z}
\right]\\
&=
-\frac{z}{2R}+\order{\frac{z^2}{R^2}}.
\label{eq:z_balance}
\end{align}
The bracket equals $\ln R+\order{1}$ as $R\to\infty$: the only unbounded term is the logarithm,
while $t_m'(z)$ and $z/(R_m+z)$ remain bounded because $t_m(z)$ is $C^1$ on a bounded interval of
$z$. Therefore
\begin{equation}
\frac{dz}{dR}\propto -\frac{z}{2R\ln(R/R_m)}.
\label{eq:z_asympt_ode}
\end{equation}
Integrating,
\begin{equation}
z(R)\propto \frac{z_\ast}{\sqrt{\ln(R/R_0)}},
\label{eq:z_asympt}
\end{equation}
where $z_\ast$ and $R_0$ are constants set by matching to the earlier evolution.

Since $t_m(z)$ is smooth near $z=0$, the term $R_m-c_0 t_m(z)$ remains bounded as $R\to\infty$.
Also,
\[
\ln\frac{R}{R_m+z}=\ln\frac{R}{R_0}+\order{1}.
\]
Equation~\eqref{eq:l_z_leading} then gives
\begin{equation}
\ell(R)=z(R)\ln\frac{R}{R_0}+\order{1},
\end{equation}
and therefore
\begin{equation}
\ell(R)\propto \sqrt{\ln\frac{R}{R_0}},
\label{eq:l_asympt}
\end{equation}
which is Eq.~\eqref{eq:1}.

\section{Video Analysis}\label{video}
\def\ntheta{n} The high-pressure front is not trivial to observe in
the footage of the Beirut explosion, as it only appears as a slightly
different hue against the sky background. However, we were able to
observe it in some footage, which we will refer to as VHP (video of
high-pressure). VHP is the clip starting at timestamp 1:46 in
Ref.~[\!\!\citenum{VHP}]. The video was first contrast-enhanced and
brightness-adjusted in the video editor Clipchamp\cite{Clipchamp} to
improve the visibility of the high-pressure front. This enhanced video
is provided in the Supplementary Material. Each frame of the selected
video was analyzed in the video analysis software Logger
Pro\cite{LoggerPro} to track both the outer edge of the visible
condensation cloud and the faint leading edge of the high-pressure
front. Frame numbers were converted to time using the video’s 30
frames-per-second rate, with $t=0$ set to be the instant of the
explosion's initial flash. We analyzed the video at times
$1.93\text{ s} < t < 3.17\text{ s}$. For each frame, the number of
pixels from the explosion epicenter to the outer edge of the
condensation cloud ($\ntheta_{R-\ell}$) and to the leading edge of the
high-pressure front ($\ntheta_{R}$) was counted. We estimated a
$\sigma_{\text{VHP}} = \pm$5 pixel uncertainty in measurements
obtained from VHP based on visibility limits and the blurriness of the
pressure front.

Pixel distances $n_r$ in the video are proportional to angular
distances. Although some lens distortion may occur near the edges
of the frame, we assume that the measurements analyzed here are
sufficiently close to the center that such effects are negligible.
Under this assumption, each pixel
has the same angular size, given by the camera's viewing angle
$\alpha$ divided by its number $N$ of  pixels in the horizontal direction, 
\begin{equation}
\theta_{\text{px}} = \frac{\alpha}{N}.
\end{equation}
The horizontal pixel resolution of the VHP clip is  $N=1920$. We take
$\alpha = 70^\circ$ ($1.22$ rad), which is a typical viewing angle for a
cellphone camera. To convert
$\ntheta_{R}$ and $\ntheta_{R-\ell}$ from pixels to radians, we multiply them by
the angular size $\theta_{\text{px}}$ of a single pixel; for example,
\begin{equation}
  \label{eq:1bis}
  \theta_R = n_R \theta_{\text{px}}.
\end{equation}
\begin{figure}[htb]
\includegraphics[scale=0.6]{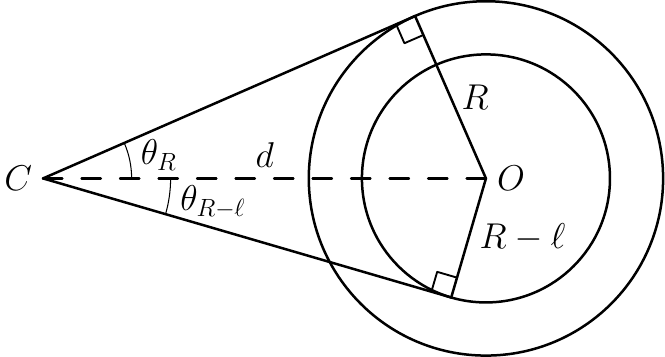}
\caption{\label{fig:Geometry} Geometry and
measured quantities in VHP. $O$, $C$ mark positions of the explosion
and of the camera. $R$ is the radius of the shock front and $R-\ell$ is
the radius of the Wilson cloud. Angles $\theta_{R}$ and $\theta_{R-\ell}$
are determined by counting pixels. From their values we obtain the
width $\ell$ of the overpressure region.}
\end{figure}
From the geometry shown in Fig.~\ref{fig:Geometry}, the shock radius
$R$ and the thickness $\ell$ of the high-pressure layer are given by
\begin{align}
 \label{eq:R}
R &= d \sin \theta_{R},
\\
 \label{eq:ell}
\ell &= d \left(\sin \theta_{R} - \sin \theta_{R-\ell} \right).
\end{align}
The distance $d$ from the camera to the explosion is determined using
independent measurements \cite{Rigby:2020}  of the speed of the shock front $D$.
Assuming that the speed of the shock front is constant, we have that
\begin{equation}
D = \frac{R \left(t_2 \right) - R \left(t_1 \right)}{\Delta t}
=\frac{d \left[ \sin \theta_R \left(t_2 \right) - \sin \theta_R \left(t_1 \right) \right]}{\Delta t},
\end{equation}
with $\Delta t = t_2 - t_1$. Solving for $d$,
\begin{equation}
 \label{eq:d}
d =\frac{D \Delta t}{\sin \theta_R \left(t_2 \right) - \sin \theta_R \left(t_1 \right)}.
\end{equation}
Taking\cite{Rigby:2020} $t_1 = 1.933$ s, $t_2 = 3.167$ s, and $D = 363\pm5$ m/s, we obtain
\begin{equation}
d \approx 3400\,\text{m}.
\end{equation}
\begin{figure}[htb]
\includegraphics[scale=0.46]{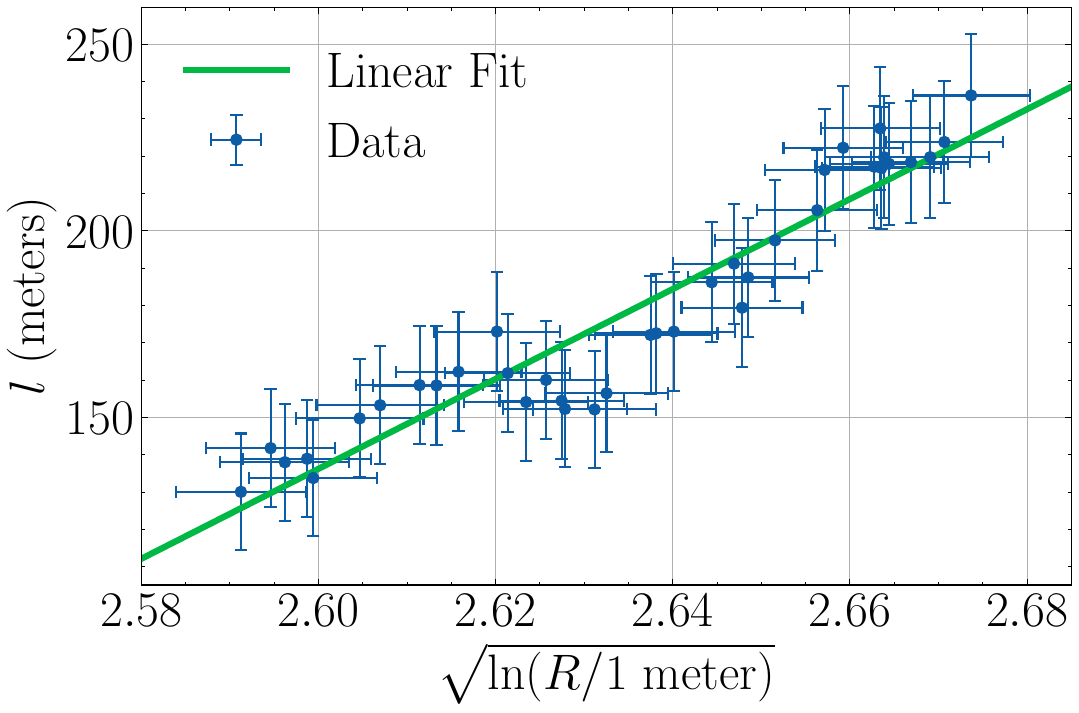}
\caption{\label{fig:Landau} Thickness $\ell$ of the high-pressure
  layer versus $\sqrt{\ln (R/1 \text{ meter})}$. Error bars are
  $1\,\sigma$, obtained by propagating a $\pm 5$-pixel picking
  uncertainty using 6$\times 10^{-4}$ rad/px scale. The solid line is
  a linear fit to $\ell=a\sqrt{\ln R}+b$, with the coefficient of
  determination of 0.9. The observed linearity is consistent with the
  scaling in Eq.~\eqref{main} for layer expansion.}
\end{figure}
With $R$ and $\ell$ determined for each frame, we examine how the
thickness of the high-pressure layer evolves with shock radius. To
compare the measurements with the prediction of Eq.~\eqref{main}, we
plot $\ell$ versus $\sqrt{\ln R}$, as shown in
Fig.~\ref{fig:Landau}. Equation~\eqref{main} predicts that $\ell$
should grow linearly with $\sqrt{\ln R}$. We therefore fit a linear
function to the data and observe a strong linear trend, with a
coefficient of determination of 0.9. This agreement supports the
applicability of Eq.~\eqref{main} to the evolution of the
high-pressure layer in a weak spherical shock.

\subsection{Uncertainty Propagation}
Here we derive the uncertainties used to plot the error bars in
Fig.~\ref{fig:Landau}. For ease of notation, we denote the uncertainty
in a variable $x$ by $\sigma_x$. We have estimated
$\sigma_{\text{VHP}} = \pm 5$ pixel uncertainty in measurements
obtained from VHP.

From Eq.~\eqref{eq:1bis}, $\sigma_{\theta_R} = \sigma_{\text{VHP}}
\theta_{\text{px}}$. This leads to $\sigma_{\sin \theta_R} = |\sigma_
{\theta_R} \cos \theta_R|$. $\sigma_{\sin \theta_{R-\ell}}$ is derived
analogously. 

To find $\sigma_d$ for $d$ in Eq.~\eqref{eq:d}, we have $\sigma_D = 5$
m/s, and $\Delta t$ has no uncertainty. Thus,  
\begin{equation}
\sigma_d = d\sqrt{\left( \frac{\sigma_D}{D}\right)^2 + \frac{2
    \sigma_{\sin \theta_R}^2}
{\left[\sin \theta_R (t_1) - \sin \theta_R (t_2) \right]^2}}.
\end{equation}
Now we can find $\sigma_R$ and $\sigma_{\ell}$ for Eqs.~\eqref{eq:R} and \eqref{eq:ell}:
\begin{align}
\sigma_R &= R \sqrt{\left( \frac{\sigma_d}{d}\right)^2 
+ \left( \frac{\sigma_{\sin \theta_R}}{\sin \theta_R}\right)^2},
\\
\sigma_{\ell} &= \ell \sqrt{\left( \frac{\sigma_d}{d}\right)^2 +
        \frac{\sigma_{\sin \theta_R}^2 + \sigma_{\sin \theta_{R-\ell}}^2}
                {\left(\sin \theta_R - \sin \theta_{R-\ell} \right)^2}
}.
\end{align}
The vertical error bar in Fig.~\ref{fig:Landau} is
$\sigma_{\ell}$. The horizontal error bar is given by 
\begin{align}
\sigma_{\ln R} &= \frac{\sigma_R}{R},
\\
\sigma_{\sqrt{\ln R}} &= \frac{\sigma_{\ln R}}{2 \sqrt{\ln R}}.
\end{align}

\subsection{Humidity Correction}\label{sec:hum}
We have assumed that the condensation cloud forms immediately after
the over-pressure layer passes, that is, when the pressure $p$ drops
below the ambient pressure $p_0$. However, cloud formation actually
begins when the relative humidity reaches $S = 1$, which occurs
slightly after  \cite{Waltz} $p = p_0$. Here we attempt to account for
this time delay.

For this discussion, we convert our measured distances and times to
units of the dynamic distance $R_0 = \left(E/p_0\right)^{1/3}$ and the
dynamic time $t_0 = E^{1/3} p_0^{-5/6} \rho_0^{1/2}$, where $E$ is the
energy of the explosion\cite{Waltz}. Here we take
$E = 2.3 \times 10^{12}$ J (from Ref.~[\!\!\citenum{Diaz2021Beirut}]),
$p_0 = 100.6$ kPa [\!\!\citenum{weather}], and $\rho_0 = 1.2$
kg/m$^3$, which gives $R_0 \approx 280$ m and $t_0 \approx 1$ s.

Ref.~[\!\!\citenum{Waltz}] expresses the relative humidity as
$S = S_0 M$, where
\begin{equation}
 \label{eq:WaltzM}
M = \frac{p}{p_0} \exp \left[ 17.45 \left(\frac{300 \text{ K}}{T_0} \right)
\left( \frac{p_0 \rho}{p \rho_0} - 1\right) \right]
\end{equation}
is the relative humidity multiplying function. Since we are in the
weak-shock regime and the adiabatic relation
$p\rho^{-\gamma}=\text{const.}$ holds, we may rewrite
Eq.~\eqref{eq:WaltzM} as
\begin{equation}
 \label{eq:WaltzMAdiabatic}
M = \frac{p}{p_0}
\exp \left[
17.45 \left(\frac{300 \text{ K}}{T_0} \right)
\left( \left(\frac{p}{p_0} \right)^{\frac{1}{\gamma}-1} - 1 \right)
\right].
\end{equation}
\begin{figure}[htb]
\includegraphics[scale=0.49]{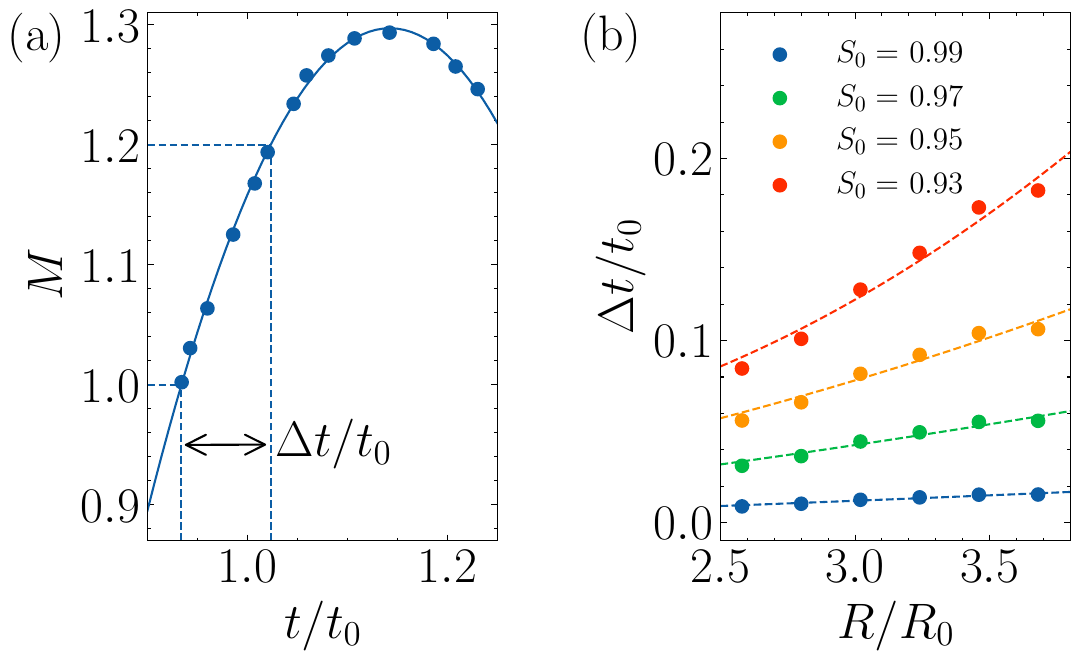}
\caption{\label{fig:DelayTime} \textbf{(a)} Example of extracting the
  time delay $\Delta t/t_0$ due to humidity for $S_0 = 0.83$
  ($M = 1.2$) from a profile of $M$ at $R/R_0 = 1.26$. The blue data
  points were extracted from Fig.~1 of Ref.~[\!\!\citenum{Waltz}],
  while the solid blue line is a parabola fitted to the extracted
  data.  \textbf{(b)} Extracted time delays $\Delta t/t_0$ at
  different $R/R_0$ for several selected values of $S_0$.  }
\end{figure}
For a particular air parcel, we seek the time delay $\Delta t/t_0$
between the time at which $p = p_0$ and the time at which $S =
1$. When $p = p_0$, we have $M = 1$, and when $S = 1$, we have
$M = S_0^{-1}$. The quantity $\Delta t/t_0$ can be extracted from the
curves in Fig.~1 of Ref.~[\!\!\!\citenum{Waltz}]. To do this, the
curves were digitized using the WebPlotDigitizer software
\cite{WebPlotDigitizer} and then fitted with
parabolas. Figure~\ref{fig:DelayTime}(a) illustrates an example of
extracting $\Delta t/t_0$ for $S_0 = 0.83$ for an air parcel at
$R/R_0 = 1.26$.

Figure~\ref{fig:DelayTime}(b) shows the extracted time delays for the
region relevant to this work ($2.5 < R/R_0 < 3.75$) for several
selected values of $S_0$. Since we observe the Wilson cloud in this
region, $M$ cannot exceed approximately $1.1$, which corresponds
\cite{Waltz} to $S_0 \approx 0.9$. Thus we consider relative
humidities in the range $0.9 < S_0 < 1$. The extracted values of
$\Delta t/t_0$ were fitted with parabolas to obtain interpolated
values at each $R/R_0$.

To correct for the time delay in our measurements of $R-\ell$, we
shift each measurement
$\left[(R-\ell)/R_0, t/t_0\right]\rightarrow \left[(R-\ell)/R_0, t/t_0
  - \Delta t/t_0\right]$.  This shift causes a misalignment in time
between our measurements of $R$ and $R-\ell$. To determine
$\ell = R - (R-\ell)$ at each time, we interpolate the measurements of
$R$ and the time-corrected values of $R-\ell$ using cubic splines and
then subtract the interpolated curves.
\begin{figure}[htb]
\includegraphics[scale=0.46]{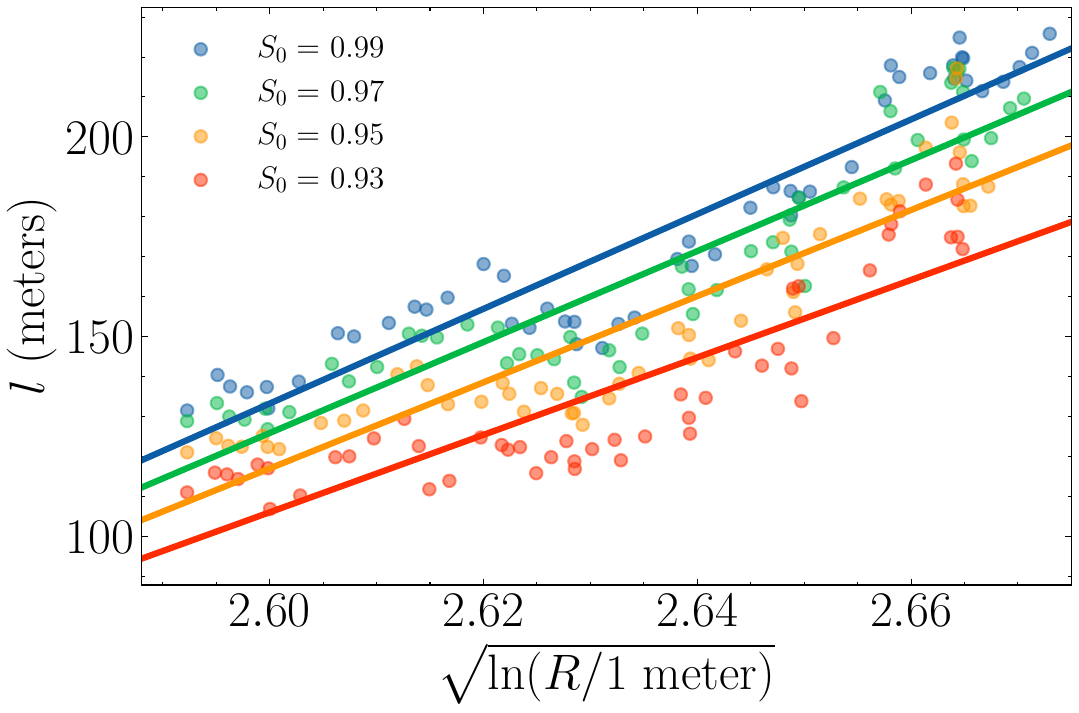}
\caption{\label{fig:HumidityCorrectedLandau} Thickness $\ell$ of the
  high-pressure layer versus $\sqrt{\ln (R/1\,\text{m})}$ for several
  values of $S_0$. The solid lines are linear fits to
  $\ell = a\sqrt{\ln R} + b$. Error bars are not included because the
  data points were obtained from interpolated curves. Lower $S_0$
  primarily results in a shift in $\ell$, while the slope does not
  change significantly.  }
\end{figure}

The measurements of $\ell$ versus $\sqrt{\ln R}$ corrected for the
time delay at several $S_0$ are shown in
Fig.~\ref{fig:HumidityCorrectedLandau}, together with linear
fits. Qualitatively, we observe that varying $S_0$ primarily shifts
the values of $\ell$, while the slope of the dependence remains nearly
unchanged. At lower $S_0$, the linear fit becomes slightly less
accurate, as indicated by a larger spread of the data points about the
fitted line. Nevertheless, we still observe a growth of $\ell$ at
lower $S_0$ as the shock wave evolves. This indicates that the
logarithmic growth given in
Eq.~\eqref{main} is robust to reasonable variations in ambient
humidity.

The full analysis of the data is provided as a Jupyter Notebook in the
Supplementary Material.

\subsection{Qualitative Observations of Wilson Cloud Dynamics}
In the video analyzed in this work, we observe the Wilson cloud expand
and eventually disappear. Ref.~[\!\!\citenum{Waltz}] provides
theoretical insight into the lifetime of the Wilson cloud, which
depends strongly on the relative humidity $S_0$. Higher values of
$S_0$ lead to longer cloud lifetimes. According to
Ref.~[\!\!\citenum{Waltz}], $S_0$ must be at least $0.7$ for any
appreciable cloud to form, and at such low values the lifetime is
expected to be very short. For example, when $S_0 = 0.75$, the cloud
disappears at approximately $t/t_0 = 1$.

The relative humidity during the Beirut explosion, however, was
reported \cite{weather} to be $S_0 = 0.73$, yet the Wilson cloud is
observed to persist until approximately $t/t_0 = 4$. We also observe
that the cloud begins to disappear from the bottom first and persists
longer at higher altitudes. We attribute this behavior to an increase
in relative humidity with altitude, which may arise from vertical
temperature variations in the atmosphere, with warmer air near the
ground and cooler air at higher elevations. Indeed, clouds are visible
near the altitude where the Wilson cloud persists longest, indicating
that the relative humidity there must be close to saturation
($S_0 \approx 1$). The measurements presented in this work were all
taken at approximately the same altitude, so we do not expect
additional time corrections due to vertical $S_0$ variations.

\section{Pedagogical aspects of the analysis}
\label{sec:pedagogy}

Use of the Beirut explosion in teaching has been proposed by Echiburu
et al.\cite{Echiburu2023}. They formulated three context-rich problems
based on images, videos, maps, and audio recordings. Emphasizing
active learning, collaboration, and discussion based on scientific
evidence, they focus on classical-mechanics topics of dimensional
analysis, energy estimation, and wave propagation. The present work is
complementary: it extends this approach to a compressible-flow
phenomenon.

The pedagogical value of the present analysis lies in challenging
students to identify structures in the recordings (overpressure layer)
and to connect them to theory.  They have to exploit imperfect data
and estimate uncertainties.

Learning goals include: to identify a meaningful observable in video
data; to relate that observable to weak-shock theory; to test the
predicted broadening law against imperfect observations; and to assess
how uncertainties affect the strength of the conclusion. A practical
classroom implementation could be guided by an instructor providing
selected frames and partial measurements. Alternatively, it can be a
longer project in which students select frames and extract data
themselves. Guiding questions include: Which visible fronts can be
measured? What assumptions are required to extract distances from
images? How should uncertainties be propagated? What effects may
explain deviations from the predicted scaling?

The same approach can be applied to follow-up projects.  Examples
include studying water waves generated by a blast above the sea
surface and the lifting of dust or debris by blast-induced flow.
These examples show that publicly available videos can support a
broader set of open-ended student investigations in fluid dynamics and
nonlinear wave phenomena.

\section{Conclusion}
Our results support the theory of shock waves far from the
explosion. We find evidence of the broadening of the overpressure
layer and a linear relationship between its thickness $\ell$ and
$\sqrt{\ln{R}}$.

We believe that this project has a significant pedagogical value. It
illustrates non-trivial but tractable theoretical considerations with
a striking phenomenon visible in published videos. Such a connection
is relatively rare, since shock waves are not frequently visible.

In our turbulent times, awareness of shock waves is valuable for
students. It fosters intuition that supports safer design of
future structures that may be subjected to shocks.
Such awareness can also alert students to dangers: in the Beirut
catastrophe, people standing near windows were injured when the glass
suddenly shattered.

In the future, related research opportunities include studying water
waves generated by shocks propagating above the sea surface, as seen
in some Beirut videos, and investigating the relationship between
missile explosions and the lifting of dust from surrounding surfaces,
as observed in many recent drone recordings. 

\begin{acknowledgments}
We are grateful to John Dewey for many helpful remarks. We thank Jan
Czarnecki for discussions of the cloud in the underpressure layer of
the blast wave and Rene Payne for collaboration at an early stage of
this project.
We used FeynDiagram\cite{fey} and Asymptote\cite{Asymptote} to draw
diagrams.
We used Chat GPT 5 for literature research,  checking text, and 
help with Asymptote. 
This research was supported by Natural Sciences and
Engineering Research Canada (NSERC).
\end{acknowledgments}

\section*{Author Declarations} 
The authors have no conflicts to disclose.
\section*{Data Availability Statement} 
The data that support the findings of this study  can be found in the
supplementary materials. 
%

\end{document}